\documentclass[12pt,a4paper]{article}

  \usepackage{a4wide}
  \usepackage{latexsym}
  \usepackage{epsf}
  \usepackage{amssymb,bbm}
  \usepackage{graphicx}
  \usepackage{amsmath, cite}
  \usepackage{amsmath,amssymb,amsthm}
  \usepackage{verbatim}
  \usepackage{hyperref}
\renewcommand{\d}{\textrm{d}}
\newcommand{\e}{\textrm{e}}

\newcommand{\w}{\wedge}

\newcommand{\be}{\begin{equation}}
\newcommand{\ee}{\end{equation}}

\begin{document}
\numberwithin{equation}{section}

\begin{center}

\begin{flushright}
{\small UUITP-22/13\qquad UG-FT 307/13 \qquad CAFPE 177/13} \normalsize
\end{flushright}
\vspace{1.5 cm}

{\LARGE {\bf BPS domain walls from \\ \vspace{0,5cm} backreacted orientifolds}}

\vspace{1 cm} {\large  J. Bl{\aa }b\"ack$^a$, B. Janssen$^b$, T. Van
Riet$^c$, B. Vercnocke$^d$}\\

\vspace{1 cm}{$^{a}$ Institutionen f{\"o}r fysik och astronomi,\\ Uppsala Universitet,\\ Box 516, SE-751 08 Uppsala, Sweden}

\vspace{0.2 cm}  \vspace{.15 cm} {$^b$ Departamento de F\'{\i}sica
Te\'orica y del Cosmos\\ and Centro Andaluz de F\'{\i}sica de
Part\'{\i}culas Elementales,\\ Universidad de Granada, 18071
Granada, Spain}

\vspace{0.2 cm}  \vspace{.15 cm} {$^c$ Instituut voor Theoretische Fysica, K.U. Leuven,\\
Celestijnenlaan 200D B-3001 Leuven, Belgium}

\vspace{0.2 cm}  \vspace{.15 cm} {$^d$ Institute of Theoretical Physics, University of Amsterdam,\\
Science Park, Postbus  94485, 1090 GL Amsterdam, Netherlands}

\vspace{0.7cm} {\small \upshape\ttfamily johan.blaback @
physics.uu.se,  bjanssen @ ugr.es, thomasvr @ itf.fys.kuleuven.be, bert.vercnocke @ uva.nl}  \\

\vspace{1.2cm}

{\bf Abstract}
\end{center}


{\small 
Compactifications with D-brane and orientifold sources lead to  standard gauged supergravity theories if the sources are smeared over the internal directions. It is therefore of interest to find how the solutions described by the gauged supergravity are altered by properly localising the sources. In this paper we analyse this for BPS domain wall solutions in the seven-dimensional gauged supergravity obtained from an O6 toroidal orientifold in massive IIA supergravity. This is one of the simplest no-scale supergravities that can be constructed and analysed in full detail. We find and discuss the BPS domain walls both when the O6 planes are smeared and localised. When the O6 planes are localised the domain wall solutions live in a warped compactification.   In order to get explicit expressions we also consider the non-compact versions of the solutions for which the O6 planes have been traded for D6 branes. Through T-duality we obtain partially localised solutions for compactifications to four dimensions using O3 planes with 3-form fluxes.} 

\setcounter{tocdepth}{2}
\newpage
\tableofcontents
\newpage

\section{Introduction}
Flux compactifications of string theory that are of phenomenological interest typically involve the use of space-filling D-brane or orientifold sources. Perhaps the most relevant reason is that such sources seem necessary for flux solutions with a separation between the Kaluza-Klein (KK) scale and the scale of the vacuum energy \cite{Tsimpis:2012tu, Petrini:2013ika}. If such a separation is absent, the vacuum does not appear as four-dimensional to an observer that can do experiments with energy at least as large as the vacuum energy. A second reason for the use of sources is their importance in de Sitter model building \cite{deWit:1986xg,Maldacena:2000mw, Townsend:2003qv}. For both the issue of scale separation and dS model building the sources should have net negative tension (orientifold planes), at least at leading order in the $\alpha'$ and $g_s$ expansion \cite{Petrini:2013ika, Danielsson:2009ff, Wrase:2010ew}. A third reason is their importance in creating sufficient warping, which in turn is useful in attempts to solve the hierarchy problem \cite{Giddings:2001yu}, or to warp down supersymmetry-breaking terms \cite{Kachru:2003aw}.  Other reasons are that D-branes are used in type II constructions of standard model building \cite{Blumenhagen:2005mu,Blumenhagen:2006ci} and they are a source of supersymmetry breaking.   

When these sources are parallel, or when they allow an F-theory understanding, the backreaction can be computed exactly, at least for supersymmetric vacua \cite{Giddings:2001yu, Grana:2006kf, Blaback:2010sj}.  Unfortunately some of the most intriguing flux solutions necessarily involve intersecting sources whose backreaction has not yet been computed. Most notable are the flux compactifications of type IIA \cite{Behrndt:2004km, Derendinger:2004jn, DeWolfe:2005uu, Caviezel:2008ik} and type IIB supergravity \cite{Caviezel:2009tu, Petrini:2013ika, Solard:2013fva} to supersymmetric $AdS_4$ vacua which have astonishing features: they allow moduli stabilisation at the classical level, an arbitrary separation of scales together with arbitrary weak coupling, small cosmological constant and large volume such that these solutions are under maximal control. Unfortunately these solutions have only been studied in the smeared limit (see \cite{Acharya:2006ne} for a discussion on that matter) and little progress has been made on how the backreaction of the localised sources could be described, apart from the investigation in \cite{Saracco:2012wc}. It has even been suggested that the backreaction can invalidate the existence of the solutions \cite{McOrist:2012yc} or be very large such that the smeared limit is far from the actual solution \cite{Banks:2006hg}.  This clearly indicates that computing the backreaction of space-filling sources in flux solutions is of crucial importance. 

There are two approaches to this problem: one can simply attempt to compute the exact backreaction. Or one could try to understand the effect of the backreaction in the low energy effective action. This last option implies one does not need to know the full backreaction, but rather needs to understand its integrated effect. These attempts go by the name of \emph{warped effective field theory} (WEFT) and a sample of papers on this topic is \cite{Shiu:2008ry, Frey:2008xw, Martucci:2009sf, Frey:2013bha}. However, several important issues that depend on local features, like those raised in \cite{Banks:2006hg, McOrist:2012yc, Saracco:2012wc}  for supersymmetric AdS vacua or the issues related to the stability of the anti-branes \cite{Blaback:2012nf} in the KKLT scenario \cite{Kachru:2003aw}, cannot be addressed by the integrated approach of WEFT.\footnote{The stability of these branes is questioned as a consequence of the singular three-form fluxes when the anti-branes backreact \cite{McGuirk:2009xx, Bena:2009xk, Bena:2012bk, Blaback:2011nz, Gautason:2013zw}.  A review on the state of the art of backreaction computations can be found in \cite{Junghans:2013xza}.} Therefore we are interested in computing the exact backreaction.

There are at least two relevant circumstances for which exact backreaction has not been computed: dynamical solutions in models with parallel sources and both vacuum and dynamical solutions in models with intersecting sources. In this paper we address the first of these. We compute the backreaction of orientifold planes in BPS domain wall solutions of no-scale supergravity theories.  Apart from the classical no-scale Minkowski vacuum one can consider solutions that involve non-constant fields. Domain walls are such a particular class of solutions that probe the structure of the scalar potential, since the scalar fields of such solutions are non-constant in space. We found it natural to treat the BPS domain walls first, due to the supersymmetry conditions.  One of our motivations to study domain walls, apart from the technical reasons mentioned above, is their interpretation as tunnelling processes in flux vacua landscapes \cite{Ceresole:2006iq}. It has been argued that the warping can be relevant in the tunnelling process  \cite{Ahlqvist:2010ki}, which was ignored in the gauged supergravity analysis of \cite{Ceresole:2006iq}. The arguments of \cite{Ahlqvist:2010ki} rely on WEFT, but it would clearly be beneficial to have the full 10-dimensional solution describing the tunnelling process. In this paper we will not verify the effect of warping on the tunnelling process but instead provide the technical tools necessary for this. 

This paper is a continuation of earlier work \cite{Blaback:2012mu}, in which we formulated a simple supergravity model and were able to compute the backreaction for a rather simple domain wall flow. This model is the simplest no-scale supergravity we are aware of and it lives in seven space-time dimensions. It can be found from a compactification of massive IIA supergravity on $\mathbb{T}^3/\mathbb{Z}_2$, with 8 space-filling O6 planes and the appropriate $H_3$ flux and Romans mass contribution $F_0$ to cancel the RR tadpole. The no-scale Minkowski vacuum breaks supersymmetry, in a way similar to the no-scale solutions in four dimensions from IIB orientifolds with ISD fluxes that contain a $(0,3)$-piece \cite{Giddings:2001yu} (in fact our model is T-dual to such a set-up). The simplest domain wall flows through moduli space\footnote{In this paper moduli space refers to the space of continuous massless deformations of the solution.} and its backreaction was computed exactly \cite{Blaback:2012mu}. In fact, it already existed in an earlier paper \cite{Janssen:1999sa}, but was left without interpretation at the time. Since the flow occurs in the moduli space, such a domain wall does not probe the potential energy. In this paper we consider all BPS domain wall flows, hence also those that move upward in the potential and compute their backreaction. Again we can rely on earlier investigations of BPS solutions in massive IIA \cite{Imamura:2001cr} that we give a new interpretation. 

The practical motivation for studying the O6 massive IIA compactification to seven dimensions is its simplicity, allowing us to classify all BPS domain walls and find their explicit warped versions. However, this simple flux compactification is interesting in its own right, since it relates to the debate on the consistency of O6 planes in massive IIA string theory \cite{McOrist:2012yc}. There is no known lift of the O6 planes in massive IIA where the singularity of the orientifold space time can be resolved. Recently, in \cite{Saracco:2012wc} an intriguing observation was made that O6 solutions in massive IIA could be regular already prior to any possible uplift. Our massive IIA Minkowski solution is simple enough to verify this idea explicitly within a compact setting.

The rest of this paper is organised as follows.  We start with section \ref{compactversusnoncompact}, where we describe a formal map between compact (O$p$) and non-compact (D$p$) solutions since the non-compact ones allow a more analytic treatment of backreaction. In section \ref{sec:smeared} we compactify massive IIA with smeared O6 planes to seven-dimensional gauged supergravity and we solve for the BPS domain walls. We compute the backreaction of the localised sources in these domain wall solutions in section \ref{sec:localised}. To that end we review all the non-compact 1/4 BPS solutions in massive IIA that possess the proper symmetries to be interpreted as domain walls and we give the corresponding compact orientifold solutions. Only a subset of these solutions describe the backreaction of the smeared domain walls. 
In section \ref{sec:4D} we initiate the same program for four-dimensional IIB orientifold compactifications, which are of more interest. By a simple T-duality we obtain the solutions corresponding to the partially backreacted four-dimensional domain walls. Section \ref{sec:discussion} contains a discussion and various avenues for new research.

\section{\texorpdfstring{D$p$/O$p$}{} sources smeared and localised}\label{compactversusnoncompact}

Solutions describing orientifold compactifications tend to be (partly) implicit because the equations of motion involve the Laplacian on the internal space. Solutions to these generalised Laplace equations are not analytically known in general. When certain problems require explicit expressions it becomes desirable to find a way around this. A common technique is to describe the solution only locally and not to worry about compactness. For example one could consider polar coordinates on a torus and make an Ansatz that only depends on the radial variable. The generalised Laplace equation can then often be solved analytically and almost always numerically. This solution will however not be complete since polar coordinates do not cover the whole torus.\footnote{This is well known for the non-compact Klebanov--Strassler conifold solution \cite{Klebanov:2000hb}, which should be a local description of a highly warped conifold point in a compact CY orientifold construction of \cite{Giddings:2001yu}.} For a class of models, including the one we discuss in this paper, there is a different technique to find analytic expressions. This relies on changing the O$p$ planes with D$p$ branes as we now explain. 

Consider the following Bianchi identity:
\begin{equation}
\d F_{8-p} = H_3\wedge F_{6-p} + \rho_p\,,
\end{equation}
where the charge distribution $\rho_p$ is given by $\rho_p=Q_p\delta_p$ for a localised O$p$/D$p$ source.  If the space $M$ transverse to the O$p$/D$p$ source is compact then we can derive the following equation by integration
\begin{equation}\label{Gauss}
Q_p= -\int_{M} H_3\wedge F_{6-p} \,.
\end{equation}
Hence, if no fluxes $H_3$ and $F_{6-p}$ are present we get the usual consequence of Gauss' law that a compact space cannot be charged. In the presence of fluxes this can be remedied by adding the exact opposite charge dissolved in fluxes. If, on the other hand, the space is non-compact, there is no such constraint. Almost all of the \emph{BPS} flux solutions in the literature are such that for D$p$ branes and O$p$ planes the following relation between the signs exists
\begin{align} 
&\text{sign}[Q_{Op}] = -\text{sign}[\int_{M} H_3\wedge F_{6-p}]\,,\label{signsOp}\\ & \text{sign}[Q_{Dp}] = +\text{sign}[\int_{M} H_3\wedge F_{6-p}]\,.\label{signsDp}
\end{align}
Therefore a generic BPS flux solution is compact when O$p$ planes are invoked and non-compact for D$p$ branes. In what follows we refer to such models as `vanilla models'. A simple intuition for (\ref{signsOp}), (\ref{signsDp}) was presented in \cite{Blaback:2010sj}. One can think of the BPS orientifold solutions as solutions that have a balance in forces between two objects:  the localised (delta-peaked) orientifold sources and the  smooth flux distribution $H\wedge F_{6-p}$ which carries opposite charge and tension. Since the tensions of sources and fluxes are opposite the gravitational repulsion between these objects is exactly counterbalanced by the electromagnetic attraction due to the opposite charges. When the O$p$ planes are replaced with D$p$-branes, which have opposite charge and tension, the same balance holds, but now gravity is attractive and electromagnetism repulsive. Because of this no-force condition the vanilla models allow a formal map between compact and non-compact solutions. For a given orientifold setup one can simply change the O$p$ charge (and tension) distribution by a D$p$ distribution.
\begin{equation}\label{map}
Op \leftrightarrow Dp\,.
\end{equation}
This interchange has the opposite sign of charge and tension and hence a compact solution will not be possible anymore if the O$p$ setup was compact as a consequence of Gauss' law (\ref{Gauss}) and (\ref{signsDp}). The formal map (\ref{map}) does not affect the bulk equations of motion, only the boundary conditions (the delta sources one has to connect to). Since compactness is lost one can hope to have some radial variable to make the generalised Laplace equations into ODE's. Perhaps the simplest example of such a correspondence  is the no-scale seven-dimensional Minkowski vacuum in massive IIA supergravity with space-filling O6 planes first described in \cite{Blaback:2010sj}. Its non-compact version coincides with an earlier supergravity solution that describes a D6 brane in massive IIA supergravity, see \cite{Janssen:1999sa }and \cite{Imamura:2001cr}. 

The main result of this paper is that the non-compact D$p$ solutions indeed carry the information of orientifold compactifications and their associated gauged supergravity description.  The domain wall flows of the gauged supergravity that follows from a \emph{smeared} orientifold compactification give rise to profiles for certain fields in ten dimensions, denoted $\Phi_{10}$, that depend on the domain wall coordinate $z$
\begin{equation}
\Phi_{10d} = \phi_{7d}(z)\,,
\end{equation}
where the expression $\phi_{7d}(z)$ is a certain combination of fields of the seven-dimensional supergravity. When we then apply the formal map (\ref{map}) and construct the localised D$p$ solution we find an expression of the form:
\begin{equation}
\Phi_{10d}(r,z) = \sum_{i=-N}^{\infty} a_i(z) r^i\,,\quad \text{where}\quad a_0(z) = \phi_{7d}(z)\,.\label{eq:phi_10}
\end{equation}
The general solution is described by a Laurent expansion in the radial coordinate $r$ of the three-dimensional transverse space and the $r$-independent term coincides exactly with the expression obtained from the seven-dimensional gauged supergravity. The coefficients in the Laurent expansion obey certain iterative relations and when combined with the requirement that the solution smears properly, all of the coefficients are defined uniquely. For compact O$p$ solutions the situation is very similar. The main difference is that the Laurent expansion is traded for a Fourier expansion on the internal torus and again the constant piece coincides with the gauged supergravity. This result was to be expected since the fields in the gauged supergravity typically correspond to the lowest Fourier mode, which is constant on the compact space. Unfortunately we have not been able to find a simple iterative set of differential equations for the Fourier coefficients. This makes the non-compact solution more attractive since the Laurent coefficients can be computed very easily.  We expect that \eqref{eq:phi_10} holds for all vanilla models. The rest of the paper is devoted to the explicit derivation of this result for the O6 compactification to $D=7$.

Before we continue we briefly want to discuss what happens to non-vanilla models. In \cite{Blaback:2010sj} a recipe for AdS solutions was given that relies on solving the RR tadpole (\ref{Gauss}) using D-brane sources instead of orientifold planes. Such solutions have a complicated backreaction \cite{Blaback:2011nz,Blaback:2011pn} with the most notable feature that the charge density in the fluxes $H_3\wedge F_{6-p}$ goes to infinity (albeit in an integrable way). The sign of this infinity violates (\ref{signsDp}) and confirms the intuitive picture; the flux cloud is attracted both magnetically and gravitationally towards the source. Recently it was found \cite{Apruzzi:2013yva} that this solution does preserve supersymmetry when $p=6$ (for other values of $p$ this is not yet investigated). This is a very interesting phenomenon and has implications for smearing and gauged supergravity \cite{Danielsson:2013qfa}: there is no seven-dimensional gauged supergravity description of the AdS$_7$ solution, regardless of whether the D6 branes are smeared or localised. Similarly, one can demonstrate that the smeared solution is not supersymmetric \cite{Danielsson:2013qfa}.

\section{Domain walls in the smeared limit} \label{sec:smeared}

In this section we give the seven-dimensional supergravity, the consistent truncation we use, and we discuss domain wall solutions of this seven-dimensional theory.

We want to stress that solutions to the seven-dimensional gauged supergravity correspond to ten-dimensional solutions with sources that are smeared in the compact directions. This is because after the dimensional reduction to a gauged supergravity we make a consistent truncation to \emph{only the lowest mass modes}. An orientifold plane source in ten dimensions that is localised in the internal dimensions has a delta-function term in the equations or motion, which can no longer hold if we keep only the lightest KK modes. For instance for the Bianchi identity of the Ramond two-form $F_2$ we have: 
\begin{equation}
 \d F_2 = F_{0}\,H_3 + \rho_6(\vec y)\,,
\end{equation}
where the charge distribution $\rho_6(\vec y)= Q_6\,\delta_{O6}(\vec y-\vec y_p)$ for an O6 source that is localised on the internal manifold with coordinates $\vec y$ at position $\vec y_p$. By restricting to only lowest mass modes of the KK tower after dimensional reduction, we are effectively truncating the KK modes to lowest order. Expanding the $\delta_{O6}$ in Fourier modes, this means that we are replacing $\delta_{O6} \to 1$ and the source is `smeared': it has a uniform charge distribution $\rho_6(\vec y) = Q_p$.

\subsection{Seven-dimensional gauged supergravity}

Consider massive IIA supergravity with $H$ flux on $\mathbb{T}^3/\mathbb{Z}_2$ and a space-filling O6 source that sits at the 8 fixed points of the $\mathbb{Z}_2$ involution:
\begin{equation}
(a, b, c)\,,\qquad \text{where}\qquad a, b, c \in \{0, 1/2 \}\,.
\end{equation} 
Dimensional reduction gives half-maximal seven-dimensional gauged supergravity without vector multiplets (vector multiplets can be added by adding D6-branes). This is a no-scale supergravity whose vacuum breaks supersymmetry. Hence the effective field theory of the fluctuation around the vacuum is not a supergravity. The seven-dimensional gauged supergravity that is obtained from direct dimensional reduction is instead capturing a consistent subset of the ten-dimensional fields that are capable of describing the spontaneous supersymmetry breaking in the vacuum.  

For the purpose of constructing solutions we only use a truncation of this theory to the seven-dimensional metric and two scalar fields, which was described in \cite{Blaback:2012mu}. For the full dimensional reduction of the bosonic sector, we refer to \cite{Danielsson:2013qfa}.

The Ansatz for the consistent reduction is given by
\begin{align}
&  \d s^2_{10} = \e^{2\alpha\varphi} \d s_7^2 + \e^{2\beta\varphi}\delta_{ab}\d y^a\d y^b\,, \\
& H_3= h\,\d y^1\wedge\d y^2\wedge\d y^3\,,\\
&  F_0 = m\,,
\end{align}
with 
\begin{equation}
\alpha=\tfrac{1}{4}\sqrt{\tfrac{3}{5}}\,,\qquad \beta=-5\alpha/3\,,
\end{equation}
and $h$ is the H-flux quantum. We denote the dilaton as $\phi$. Tadpole cancellation requires us to take the tension of the O6-plane to be\footnote{Note that $hm$ is positive.}
\begin{equation}
T = -2 h m\,.
\end{equation}
The only contribution to the scalar potential in seven dimensions comes from the Romans mass term $m$, the amount of $H$-flux denoted $h$ and the orientifold tension, combining into \cite{Danielsson:2013qfa} 
\begin{equation}
V(\phi, \varphi) = \tfrac{1}{2}\Bigl( h\e^{-\tfrac{1}{2}\phi + 6\alpha\varphi} - m\e^{\tfrac{5}{4}\phi+\alpha\varphi}\Bigr)^2\,.
\end{equation}
The action describing this two-scalar truncation of half-maximal seven-dimensional gauged supergravity is
\begin{equation}
S = \int \Bigl( \star \mathcal{R}-\tfrac{1}{2}\star\d\phi \wedge\d\phi -\tfrac{1}{2}\star\d\varphi \wedge\d\varphi  -\star V(\phi, \varphi)\label{7d_2calar_truncation}\Bigr)\,.
\end{equation}
One can demonstrate that the other scalars in the theory decouple; they are free fields that do not enter the scalar potential (see \cite{Danielsson:2013qfa}). Since these free scalar fields do not enter the superpotential for supersymmetric solutions, they are constant.\footnote{In fact, this also holds for the fake supersymmetric domain walls  defined in section \ref{ss:BPS_fakeBPS}.} Hence our truncation is not a restriction and we can classify all BPS solutions.

\subsection{BPS and fake BPS domain walls}\label{ss:BPS_fakeBPS}
 
Let us first recall the concept of fake supersymmetry for domain wall
solutions \cite{Freedman:2003ax}. Consider a general action of $D$-dimensional gravity coupled to scalars of the form
\begin{equation}\label{Bergshoeffconvention1}
S = \int \left( \star \mathcal{R}-\tfrac{1}{2}G_{ij}(\phi)\star\d\phi^i \wedge\d\phi^j-\star V(\phi)\right)\,.
\end{equation}
Domain wall solutions in $D$ dimensions have Poincar\'e invariance in $D-1$ dimensions and depend on the transversal direction only. They are described by a metric of the form
\begin{equation}\label{domainwallAnsatz}
\d s^2_D = f^2(z)\d z^2 + g(z)^2\eta_{\mu\nu}\d x^{\mu}\d x^{\nu}\,,
\end{equation}
where $\eta$ is the metric on Minkowski space. The scalar fields only depend on the $z$-coordinate. A certain domain wall solution, given by $\phi^i(z), f(z), g(z)$ is
called fake supersymmetric if there exists a function $W(\phi)$ on
the scalar target space such that the solution obeys
\begin{align}
& \dot{\phi}^i = - f\, G^{ij}\partial _j W\,,\label{flow1}\\
& \frac{\dot{g}}{g} = \tfrac{1}{2(D-2)} \,f\,W\,.\label{flow2}
\end{align}
The domain wall solution is given by a first-order gradient flow set by the fake superpotential $W$.
Consistency with the second-order equations of motion implies a
relation between the fake superpotential $W$ and the potential $V$:
\begin{equation}\label{WV}
V(\phi) = \tfrac{1}{2}G^{ij}\partial_i W\partial_j W -
\tfrac{D-1}{4(D-2)}W^2\,.
\end{equation}
Genuine supersymmetric solutions obey the above constraint for a
very specific function $W$ which is the `true' superpotential that
appears in the supersymmetry transformations of the supergravity theory. Any
other function $W$ that obeys (\ref{WV}) generates a solution of the
second-order equations of motion through first-order flow
equations. One can prove that not all solutions to the second-order
equations of motion are fake supersymmetric, although a set of local first-order Hamilton-Jacobi equations always exist.  Recently, it has been clarified
when solutions are fake supersymmetric or not \cite{Trigiante:2012eb}.

Now we go back to our model in $D=7$, with action \eqref{7d_2calar_truncation}. To write down the solutions it is convenient to consider the following base rotation to new scalars $x$ and $u$ \cite{Blaback:2012mu}:
\begin{align}
 \phi & = -\tfrac{\sqrt{15}}{8}x +\tfrac{7}{8}u \,,\nonumber\\
 \varphi & = - \tfrac{7}{8}x -\tfrac{\sqrt{15}}{8}u\,.
\end{align}
One can then construct a one-dimensional family of superpotentials as follows
\begin{equation}\label{FakeSuperpotential}
W(x, u) = \exp(-\sqrt{\tfrac{3}{5}}x) \Bigr[
|h|\exp(-u) + |m|\exp(+u) - c \Bigl]\,,
\end{equation}
with $c$ a constant. The true superpotential giving supersymmetric solutions has $c  = 0$ \cite{Blaback:2012mu}.
All other values for $c$ correspond to fake superpotentials; for $c \neq 0$ the corresponding domain wall flow is not supersymmetric.

From here on we use the gauge 
\begin{equation}\label{gauge}
f(z)=g^{-4}(z)\,.
\end{equation}
A combination of the $\dot{x}$ equation with the $\dot{g}$ equation allows to solve $g$ in terms of $x$
\begin{equation}
e^x = g^{2\sqrt{15}}.
\end{equation}
The remaining $\dot{x}$ and $\dot{u}$ equations can be simplified with the Ansatz,
\begin{equation}
e^x = (h_1 h_2)^{\tfrac{1}{2}\sqrt{\tfrac{3}{5}}}\,,\qquad e^u = (h_1/h_2)^{1/2}\,,
\end{equation}
where $h_1$ and $h_2$ obey the following first order equations:
\begin{align}
\dot{h}_1+ c \sqrt{h_1/h_2} &=2 |h| \,,\\
\dot{h}_2+ c \sqrt{h_2/h_1} &=2 |m|\,.
\end{align}

When $c=0$ we find the supersymmetric solution
\begin{equation}
h_1 = 2 |h| z + \ell_1^2,\quad h_2 = 2 |m| z + \ell_2^2\,,\label{BPS_solution}
\end{equation}
where $\ell_1^2,\ell_2^2$ are positive integration constants.
These solutions were constructed in \cite{Bergshoeff:2004nq}. For the restricted set of solutions with $|m| \ell_1^2 =|h| \ell_2^2$, the warped extension of this solution was found in our previous paper \cite{Blaback:2012mu}. In the following section we discuss the warped extension of generic BPS solutions with $|m| \ell_1^2 \neq |h| \ell_2^2$.

We have not been able to solve analytically for the general fake supersymmetric solutions\footnote{By that we mean an expressions in terms of well known and simple functions.} but it is possible to find numerical solutions. There is however a simple class of solutions that can be found explicitly for non-zero $c$. These solutions have constant scalar $u$ with $\exp(2u)=\frac{|h|}{|m|}$ and only the scalar $x$ is flowing. They are determined by
\begin{equation}\label{modulispaceflow}
h_1(z)= \Bigl(2|h|-c\sqrt{\frac{|h|}{|m|}}\Bigr)z + \ell^2\,,\qquad h_2(z) = \frac{|m|}{|h|}h_1(z)\,.
\end{equation}
Such solutions correspond to a flow through the minimum of the no-scale potential where the on-shell value of the scalar potential vanishes. When we tune the velocity of the scalar $x$ such that  $c=0$, this is the BPS flow  \eqref{BPS_solution} with $|m| \ell_1^2 =|h| \ell_2^2$. For other values of $c$ it is fake supersymmetric.

\section{Full backreaction in ten-dimensional supergravity}\label{sec:localised}

In this section we construct the warped counterpart of the domain wall solutions of the previous section. The domain walls preserve half of the supersymmetry of the half-maximal gauged supergravity. Hence one is naturally lead to investigate 1/4 BPS solutions in massive IIA supergravity that have the proper symmetries to correspond to a domain wall on the worldvolume of the O6-plane. Since there is a map between compact O6 solutions and non-compact D6 solutions as explained in section \ref{compactversusnoncompact},  we can rely on the extensive literature of BPS brane solutions in 10D supergravity. It turns out that the analysis performed in \cite{Imamura:2001cr} is the key to understand the backreaction.  
We start by reviewing the quarter BPS solution of \cite{Imamura:2001cr} in massive IIA supergravity in subsection \ref{quarterBPS}. In subsection \ref{smear} we explicitly smear these BPS solutions and compare them to the uplift of the BPS domain walls in 7D gauged supergravity. This allows us to uniquely define their backreaction in subsection \ref{unique}.

\subsection{1/4 BPS solutions in massive IIA supergravity}\label{quarterBPS}

Interestingly a general analysis of the BPS equations for an Ansatz that describes domain walls inside the D6 worlvolume in massive IIA has been carried out earlier by Imamura in \cite{Imamura:2001cr}, where an expression for the general solution was shown to be of the following elegant form in string frame:
\begin{align}
&\d s^2 = S^{-1/2} \eta_{ij} \d x^i \d x^j + K S^{-1/2} \d\tilde z^2 + K S^{1/2} \d s^2_3\,,\nonumber\\
& \e^\phi = g_s K^{1/2} S^{-3/4}\,,\nonumber\\
&F_2  = -\frac{1}{g_s} \star_3 \d_3 S\,,\nonumber\\
&H_3=  \frac{\partial }{\partial \tilde z} (K S)\star_3 1-  \d\tilde z \wedge \star_3\d _3 K \,, \nonumber\\
& F_0 = m\,,\label{Ima}
\end{align}
where $\d s_3^2$ is the metric on $\mathbb{R}^3$;  $\star_3$ and $\d_3$ are the corresponding hodge star and differential. 
In all ten-dimensional solutions, we follow the conventions and equations of motion of massive IIA supergravity as in \cite{Blaback:2011nz}.
We have used the coordinate $\tilde z$, because it is not necessarily the previous domain wall coordinate $z$ introduced in \eqref{domainwallAnsatz}. The relation will be established below.
The Ansatz is given in terms of two functions $S(r,\tilde z)$ and $K(r,\tilde z)$ that are determined by the differential equations
\begin{align}
&\nabla_3^2 S + \frac 12 \frac{\partial^2 S^2}{\partial \tilde z^2}  =  -g_sQ_6 \delta \,,\label{compactLaplace}\\
&m g_s K = \frac{\partial S}{\partial \tilde z}\,,\label{eq:Imamura_diff_eq2}
\end{align}
with $\delta$ describing the localised source on $\mathbb{R}^3$. Note that this Ansatz is written formally as a D6-NS5 solution. The function $S$ appears with the correct factors for a D6 harmonic function, $K$ for an NS5 harmonic function. 

\paragraph{Non-compact D-brane solutions.}
Let us consider two known solutions to the non-compact equations. The first is the ordinary D6 brane, which has only $r$-dependence
\be
S = 1 + \frac{g_s Q_6}{r}\,.
\ee
As is well known, this solution preserves 16  supercharges rather than 8.

A second class of solutions has a linear $\tilde{z}$-dependence in $S$. This class is parameterised by two constants $g_s$ and $\alpha$, where $\alpha$ is the coefficient of the linear piece
\begin{equation}\label{JMO}
S =  1 + \frac{g_s Q_6}{r} - \frac{\alpha^2 r^2}{6} + \alpha \tilde{z}\,, \qquad K =\frac{\alpha}{mg_s}\,.
\end{equation}
This is the BPS solution of \cite{Janssen:1999sa} where the following parameters were chosen: $g_s=1$, $\alpha=m$.

\paragraph{Compact orientifold solutions.}
We obtain the compact $O6$ solution from the above non-compact solution by taking the transverse space to be $\mathbb{T}^3/\mathbb{Z}_2$.  The compact solution has the usual caveat that the Laplacian-type equation as \eqref{compactLaplace} on a compact space may not be explicitly solvable.  

For the reader  worried about this quick jump from non-compact to compact solutions we now prove that the tadpole condition is obeyed. Consider the part of the $H_3$ field that sits inside the transversal space:
\begin{equation}
H_{\rm internal} =  \frac{1}{mg_s} \frac{1}{2} \frac{\partial^2 }{\partial \tilde z^2} S^2 \star_3 1\,.
\end{equation}
We used the definition of $K$ (\ref{eq:Imamura_diff_eq2}) to rewrite the first term for the expression of $H$ in (\ref{Ima}).
Now we rewrite this once more using (\ref{compactLaplace})
such that we end up with
\begin{equation}
H_{\rm internal} = \frac{1}{mg_s} (-g_s Q_6 \delta - \nabla_3^2 S ) \star_1 \,.
\end{equation}
When we integrate $m\,H_3$ over the internal space, we have
\begin{equation}
\int_{\mathbb{T}^3/\mathbb{Z}_2} m \,H_{\rm internal}=  -Q_6  -  \frac{1}{g_s} \int_{\mathbb{T}^3/\mathbb{Z}_2} \nabla^2 S \,.
\end{equation}
On a compact space the last term must be zero. Hence we automatically satisfy the tadpole condition (\ref{Gauss}) with Romans mass $F_0=m$.

\subsection{Smeared limit in 10 dimensions}\label{smear}

The smeared limit (`unwarped' limit) we perform is a uniform smearing over the 3-dimen\-sional space. On the level of the ten-dimensional equations of motion, one starts by replacing the delta-functions in the source terms by $1$. This affects the Einstein equations, dilaton equation and the $F_2$ Bianchi identity. Solutions to these equations will now be such that nothing depends on the internal coordinates. Finally one replaces the O6 charge that multiplies the delta function by $Q_6 = -hm$ (charge dissolved in flux), through the tadpole condition. One can easily show that the BPS equations (\ref{compactLaplace}) and (\ref{eq:Imamura_diff_eq2}) solve these equations if also in the BPS equation (\ref{compactLaplace}) the delta function is replaced by $1$:
\be
\frac 12 \frac{\partial^2 S^2}{\partial \tilde z^2} = g_s hm\,,\qquad m g_sK = \frac{\partial S}{\partial \tilde z}\,,\label{SmearedIma}
\ee
where the $H_3$-flux is written as $H_3 = h\,{\rm vol}_3$. As explained in the beginning of section \ref{sec:smeared}, the ten-dimensional uplift of solutions to seven-dimensional supergravity should correspond to such a smeared solution.

We now verify that indeed the general 1/4 BPS solution to the ten-dimensional equations of motion after smearing coincides with the uplift of the BPS solutions of seven-dimensional supergravity determined by \eqref{BPS_solution}. The latter are given by 
\begin{align}
&\d s^2 =  h_2^{-\tfrac{1}{4}}\eta_{ij} \d x^i \d x^j +  h_1^{-\tfrac{1}{2}}h_2^{-\tfrac{3}{4}}\d \tilde z^2 +  h_1^{\tfrac{1}{2}}h_2^{-\tfrac{1}{4}}\delta_{ab}\d y^a\d y^b\,,\nonumber\\
& \e^\phi = h_1^{\tfrac{1}{4}}h_2^{-\tfrac{5}{8}} \,,\nonumber\\
&H_3 =  h \d y^1\wedge \d y^2\wedge \d y^3\,,
\end{align}
and $F_2=0$. Again we have chosen the gauge (\ref{gauge}). 
Comparing to the Ansatz (\ref{Ima}), we identify:
\be
S = g_s h_2^{\tfrac{1}{2}}\,,\qquad K = g_s^{-1/2} h_1^{\tfrac{1}{2}}h_2^{-\tfrac{1}{2}}\,,
\ee
and we have the following relation between the coordinates
\be
h_1^{-1/2}\d z = g_s^{-1/2}  \d\tilde{z}\,,\label{z_coordinates}
\ee
which can be integrated to $\tilde{z} = g_s^{1/2}\,|h|^{-1}\,\sqrt{2|h|z +\ell_1^2}$. One then readily verifies that $S$ and $K$ solve equations (\ref{SmearedIma}). This establishes the consistency of our smearing procedure.

\subsection{Warped domain walls}\label{unique}

We now discuss the solution with O6 sources localised on the three-dimensional space.

\subsubsection{Warping for a simple class of domain walls} 
One explicit backreacted solution is known and is given by equation (\ref{JMO}).  This solution is to be interpreted as a solution for which the scalar fields flow through the moduli space of the no-scale Minkowski vacuum \cite{Blaback:2012mu}. This means that the corresponding domain wall solutions in the seven-dimensional gauged supergravity are the ones that have $ m\ell_1^2 = h\ell_2^2 $, since those solutions have $V=0$ throughout their flow (see discussion around (\ref{modulispaceflow})).

\begin{figure}[ht!]
\begin{center}
\includegraphics[scale=.6]{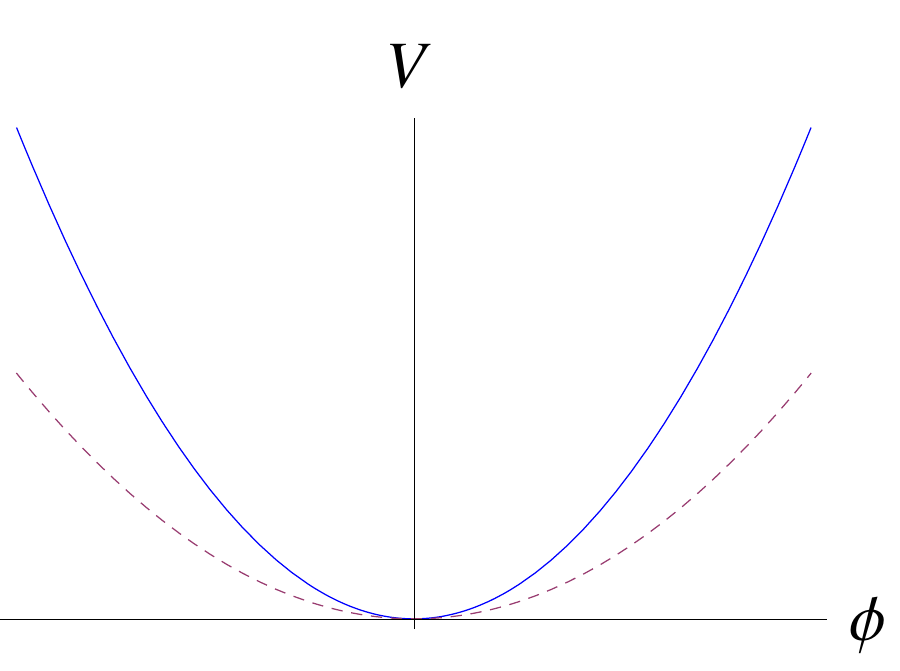} \caption{{\small \emph{The scalar potential in the gauged supergravity is depicted with the full line. The dashed line represents the potential in the warped effective field theory.}}}\label{warpedV}
\end{center}
\end{figure}

The linear $\tilde z$ dependence is not altered by warping and this is a simple consequence of the warped effective potential and the gauged supergravity potential  both keeping the same minimal value for the on-shell scalar flow \cite{Douglas:2007tu}. The rough behaviour is shown in figure \ref{warpedV}.
This viewpoint furthermore explains why ordinary D6 branes, without Romans mass and $H_3$ flux, allow a (non-supersymmetric) extension for which a piece is added to the harmonic that is linear in the worldvolume coordinates (the so-called dynamical branes\cite{Binetruy:2007tu, Gibbons:2005rt})
\be
H(r) =1 + \frac{Q_6}{r} + \sum_{i=0}^6 c_i x^i\,.
\ee 
When interpreted as a `compactification' without flux, there is no scalar potential. The linear dependence comes in through the profile of a free scalar on the worldvolume. It is the same as for solutions with flux that have a flow is in the moduli space of the vacuum.

\subsubsection{Warping for all supersymmetric domain walls} 

Using the map of orientifold solutions to D-branes, we treat the non-compact D-brane solution first. We consider a D6-brane source at the origin of  the three-dimensional transverse space, such that the solution only depends on the three-dimensional radius $r$ and of course also on the domain wall coordinate $\tilde z$. A useful suggestion for tackling this problem has been made in \cite{Imamura:2001cr}. We write the solution as a Laurent series with leading $1/r$ singular piece
\be
S(r,\tilde z) = \sum_{n=-1}^\infty a_n(\tilde z) r^n\,.
\ee
The rationale for this specific Laurent series sum is that the $1/r$ contribution gives the delta-function pole in the $S$-equation; the higher order terms cannot be excluded a priori. We do not expect any higher order $1/r^k$ poles. This Ansatz, when plugged into the equations of motion, gives a \emph{recursive} relation for the $\tilde z$ dependent coefficients \cite{Imamura:2001cr}
\be
n(n+1) a_n = -\frac 12 \frac{\partial^2}{\partial \tilde z^2}\sum_{k=0}^n a_{k-1} a_{n-k-1}\,.\label{eq:sumA}
\ee
We can solve this series by fixing the first two terms $a_{-1},a_0$. All the higher order terms are determined recursively. 
The sum \eqref{eq:sumA} for $n=0$ gives $\partial_{\tilde z}^2 a_{-1}=0$ and leaves $a_{0}$ free.  We fix these through the appropriate limits. The first term is fixed by the delta-function pole to be $\tilde{z}$ independent since it has to equal the O6/D6 charge:
\be
a_{-1} = g_s Q_6\,.
\ee
The second term can be constrained by noting that it must be such that upon smearing, the $H_3$ flux becomes $\tilde{z}$ independent and is given by the flux quantum we dubbed $h$ before: $H_3 = h\,\star_3 1$. This implies $S(r,\tilde{z}) \rightarrow S(\tilde{z}) = a_0(\tilde{z})$ and
\begin{equation}\label{a_0equation}
\frac 12 \frac{\partial^2}{\partial \tilde{z}^2}a_0^2 = g_s mh\,.
\end{equation}  
By shifting the origin of the $\tilde z$-axis, the most general solution is
\begin{equation}
 a_0= \sqrt{g_smh \tilde z^2 + \beta}\,,
\end{equation}
with one integration constant $\beta$. By comparison to the smeared limit \eqref{z_coordinates}, we see that the integration constant
 is 
\begin{equation}
\beta = \frac{g_s^2}{h}\left(h\ell_2^2 - m\ell_1^2\right)\,,
\end{equation}
and it denotes the ``distance'' from the backreacted solutions with $m \ell_1 = h\ell_2$ we discussed in \cite{Blaback:2012mu}, see eq.\ \eqref{JMO}.

For the backreaction with most general $r$-dependence, we need the other terms $a_n$. They are fixed by recursion and this uniquely defines the backreaction of the domain wall solutions of seven-dimensional supergravity.  The first terms are: 
\begin{align}
 a_1 &= -\beta \frac{g_s  mh  Q_6}{2 a_0^3}\,,\nonumber\\
 a_2   &=-\frac{1}{6} g_s mh+\beta (g_s^2  mh  Q_6)^2\left(\frac{1}{a_0^5}-\frac{5 \beta}{4 a_0^7}\right)\,,\nonumber\\
 a_3  &= \beta (g_s  mh)^2  (g_s Q_6)\left(\frac{1}{4 a_0^4}-\frac{\beta}{3 a_0^6}\right)
 +\beta(g_s^2  mh  Q_6)^3\left(-\frac{5}{2 a_0^7}+\frac{35 \beta}{4 a_0^9}-\frac{105 \beta^2}{16 a_0^{11}}\right)\,.
\end{align}
When $\beta=0$, the  $a_n$ series terminates at order 2, the only non-trivial extra term is $a_2 = -g_s mh/6$. Then we find the backreacted solution with $m \ell_2^2 = h \ell_1^2$ described in (\ref{JMO}). However, for the most general solution with $\beta\neq 0$ or equivalently $m \ell_2^2 = h \ell_1^2$, the series is infinite.

For the compact solution on $T^3/\mathbb{Z}_2$ with orientifold sources, the Laurent expansion should be replaced by a Fourier expansion
\begin{equation}
S = \sum_{\vec{n}}a_{\vec{n}}(\tilde z)\exp[i\vec{n}\cdot \vec{\theta}]\,,
\end{equation}
where $\vec{\theta}$ are the angles of the 3-torus. Assuming reality of the function $S$ and proper parity under the orientifold involution restricts the choice of $\tilde z$-dependent Fourier coefficients $a_{n}$, but we were unable to find a simple iterative equation for them. This is an interesting unsolved problem, which, if solvable, would lead to explicit solutions for KK modes.

\section{A 4D example from O3 planes}\label{sec:4D}
The seven-dimensional solutions studied above are mostly a toy model for the situation of genuine interest: flux compactifications to $D=4$. However, by simply T-dualising the seven-dimensional model we obtain a four-dimensional model. 

The T-dualised model describes an O3 compactification of IIB supergravity on a $\mathbb{T}^6$(or an orbifold thereof) to four dimensions. The three-form fluxes take a simple form. We fill one three-torus inside $\mathbb{T}^6$ with $H$-flux and the orthogonal three-torus with $F_3$ flux. This is perhaps the simplest compactification of the kind discussed in \cite{Giddings:2001yu}. Fluxes of this kind necessary lead to a non-BPS Minkowski vacuum  \cite{Blaback:2012mu}. The BPS domain wall solutions of the 4D supergravity were constructed in \cite{Blaback:2012mu} and we briefly recall those solutions before discussing the warped extension that arises when the O3 plane is localised.

\subsection{Smeared O3 compactification} 
The discussion parallels that of section \ref{sec:smeared}. However, now we need to consider a truncation of the compactification to three scalars: the dilaton $\phi$, the volume
modulus $v$ of the internal six-torus and
the volume modulus of the 3-torus threaded by the $F_{3}$-flux, called
$\chi$. The volume of the orthogonal three-torus, filled with $H_3$ depends on $v$ and $\chi$. The Ansatz, in 10-dimensional Einstein frame, is
given by
\begin{align}
& \d s^2_{10} = \exp(2\alpha v)\d s_{4}^2 + \exp(2\beta
v)\Bigl\{\exp(\gamma \chi)\delta_{ab}\d y^a \d y^b  +\exp(-\gamma\chi) \delta_{ij}\d y^i \d y^j \Bigr\}\,,\label{metric4Dreduction}\\
& H_3 = h\,\d y^1 \wedge \d y^2 \wedge \d y^3 \,,\\
& F_3 = m\,\d y^4 \wedge \d y^5 \wedge \d y^6 \,.
\end{align}
We used the following notation: $m$ and $h$ are flux quanta and they are constrained by the tadpole condition as $
T_{O3}=-hm$, the indices $a, b$ run from $1$ to $3$ and $i,j$ from $4$ to $6$.  The numbers $\alpha, \beta, \gamma$ are chosen such that we
end up in lower-dimensional Einstein frame with canonically
normalised fields:
\begin{equation}
\beta =-\tfrac{1}{3}\alpha \,,\quad \alpha^2
=\tfrac{3}{16}\,,\quad \gamma^2 = \tfrac{1}{3}\,.
\end{equation}
The lower-dimensional action is obtained from a direct dimensional reduction
\begin{equation}\label{noscalemodel}
S=\int\sqrt{-g}\Bigl(R-\tfrac{1}{2}(\partial\phi)^2
-\tfrac{1}{2}(\partial\chi)^2-\tfrac{1}{2}(\partial v)^2 - V(\phi,
v,\chi)\Bigr)\,.
\end{equation}
The scalar potential $V$ gets contributions from the fluxes and the
negative orientifold plane tension and has the form of an exact
square (due to the tadpole condition). To write it down in a clean
way we perform the following $SO(3)$ field rotation $(\phi, v,
\chi)\rightarrow (x, u, w)$:
\begin{align}
& \phi = \tfrac{1}{2}\,\,u + \tfrac{\sqrt{3}}{2}\,\,w \,,\\
& v = -x\,,\\
& \chi = - \tfrac{\sqrt{3}}{2}\,\,u + \tfrac{1}{2} \,\,w \,.
\end{align}
The scalar potential then becomes \cite{Blaback:2012mu}
\begin{equation}\label{potential}
V(x, u, w) = \tfrac{1}{2}\exp(-\sqrt{3}\,\,x) \,\,
   \Bigr[|h|\exp(-u) -
|m|\exp(+u)\Bigl]^2 \,,
\end{equation}
The scalar
potential  only depends on two scalars  ($u$ and
$x$). T-duality along  $y^4, y^5, y^6$ reproduces the O6 compactification of massive IIA.

\subsection{BPS domain walls from smeared O3 planes}
The real superpotential that determines the BPS domain wall solutions is given by
\begin{equation}\label{superpot4D}
W = \exp(-\sqrt{3}x/2) \Bigr[|h|\exp(-u) + |m|\exp(+u)\Bigl]\,.
\end{equation}
We present the solutions in the following metric gauge
\begin{equation}\label{domainwallAnsatz2}
\d s^2_4 = g^{-2}(z)\d z^2 + g(z)^2\eta_{\mu\nu}\d x^{\mu}\d x^{\nu}\,,
\end{equation}
where $\eta$ is the metric on 3d Minkowski space. We find \cite{Blaback:2012mu}:
\begin{align}
& g = (h_1h_2)^{\frac{1}{8}}\,, \label{susy1}\\
& \exp( x)= (h_1h_2)^{\sqrt{3}/4}\,, \label{susy2}\\
& \exp(u) = (\frac{h_1}{h_2})^{1/2}\,. \label{susy3}
\end{align}
where, as before,  
\begin{equation}
h_1 = 2|h| z + \ell^2_1\,,\qquad h_2= 2|m|z +
\ell^2_2 \,.
\end{equation}

\subsection{Warped solutions from localised O3 planes}

By T-dualising the 1/4 BPS solutions with lcoalised D$6$/O$6$ sources we obtain warped solutions in four-dimensions with D$3$/O$3$ sources that are localised in three directions, but smeared over the three T-duality directions $y^4,y^5,y^6$:
\begin{equation}
\begin{split}
\d s^2 &= S^{-1/2} \eta_{\mu\nu} \d x^\mu \d x^\nu + K S^{-1/2} \d \tilde{z}^2 + K S^{1/2} \d s_3^2 + S^{1/2} \delta_{ij}\d y^i\d y^j\,,\\
\e^\phi &= g_s K^{1/2}\,,\\
F_5 &= -\frac{1}{g_s} \star_3 \d_3 S \w \d y_4 \w \d y_5 \w \d y_6\,,\\
H_3 &=  \frac{\partial }{\partial \tilde z} (K S)\star_3 1- \d\tilde z \wedge \star_3\d_3 K \,,\\
F_3 &= m\ \d y_4 \w \d y_5 \w \d y_6\,.
\end{split}
\end{equation}
Note that the dilaton does not depend on $S$.

We expect the domain wall solutions with general ISD fluxes to be much more involved. A more formal method, as in \cite{Haack:2009jg}, might be useful for finding them.

\section{Discussion}\label{sec:discussion}
In this paper we explored domain wall solutions in no-scale gauged supergravities that have an interpretation in terms of orientifold compactifications of type II supergravity. The gauged supergravity does not account for the full backreaction of the orientifold planes. Accordingly, the domain wall solutions solve the ten-dimensional equations of motion only when the orientifolds are smeared over the extra dimensions. We have computed the corrections to the domain wall solutions, from a ten-dimensional point of view, that arise from proper localisation of the sources. In order to find explicit expressions we rely on a formal map between compact orientifold solutions and non-compact D brane solutions. This allows us to benefit from existing results on BPS brane solutions with fluxes, such as those in \cite{Janssen:1999sa, Imamura:2001cr}. Our analysis gives an interpretation to part of the BPS brane solutions of \cite{Janssen:1999sa, Imamura:2001cr} as describing domain walls inside the worldvolume of D-branes. In order to be as explicit as possible, we first considered a very simple no-scale orientifold compactification: the O6 reduction of massive IIA to $D=7$ \cite{Blaback:2010sj, Blaback:2012mu}. In section \ref{sec:4D} we have demonstrated how a formal T-duality relates this to the more interesting  IIB toroidal orientifold compactifications to $D=4$. The general IIB toroidal orientifold would be interesting to work out in detail. There are many other directions that this work naturally leads to and we list them here.

It would be interesting to find non-supersymmetric solutions and whether backreaction can even be computed in this case. We have reasons to believe that for fake BPS solutions this should be possible. In fact one example already exist and was presented in \cite{Janssen:1999sa}. These are simple supersymmetry-breaking domain walls that flow through moduli space and it is possible to match them to fake BPS domain walls of the gauged supergravity.

The research avenues we address here fit into a bigger question: what are the backreacted versions of \emph{all} gauged supergravity solutions? We can think of cosmological solutions, black holes and so on. In practice this means understanding the zoo of configurations inside the worldvolume of branes from a 10-dimensional point of view.

One could also wonder whether our computations could help in understanding warped effective field theory. The difficulty of formulating WEFT is performing the integration of the action over the extra dimensions and trying to identify the lower-dimensional degrees of freedom. With our approach we have the explicit 10D BPS equations and perhaps the integrated versions of these equations give the BPS equations of the WEFT, leading to the action.

\subsection*{Acknowledgements}
We would like to thank Iosif Bena, Gary Shiu and Timm Wrase for useful discussions.
TVR is supported by a Pegasus fellowship and by the Odysseus programme of the FWO.   The work of JB is supported by the Swedish Research Council (VR), and the G\"oran Gustafsson Found. The work of BJ was partially supported by the M.E.C. under contract FIS2010-17395 and by the Junta de Andaluc\'{\i}a groups P07-FQM 03048 and FQM-6552. BV acknowledges support from the European Union through the Marie Curie Intra-European fellowship 328652--QM--sing and would like to thank Evelien Dejonghe. For the initial work on this project, BV was supported by the ERC Starting Independent Researcher Grant 240210--String--QCD--BH.

\appendix

\section{\texorpdfstring{$\mathcal{N}=1$}{} supergravity formulation}

In what follows we derive the real superpotential in four dimensions from the standard $\mathcal{N}=1$ supergravity formulation for IIB orientifold compactfications with 3-form fluxes \cite{Giddings:2001yu}. Such a formulation gives a complex superpotential $\mathcal{W}$ and a real K\"ahler potential $\mathcal{K}$. The usual convention found in the literature is
\begin{equation}
S = \int \tfrac{1}{2}\star R - \mathcal{K}_{i\bar{j}}\star\d\Phi^i\wedge\d\bar{\Phi}^j - \star V \,,
\end{equation}
where
\begin{align}
& \mathcal{K}_{i\bar{j}}= \partial_i\partial_{\bar j} \mathcal{K}\,, & V = \e^{\mathcal{K}}\Bigl(D_i \mathcal{W} \overline{D_j \mathcal{W}} \mathcal{K}^{i\bar{j}} - 3 |W|^2 \Bigr)\,, \\ 
&D_i \mathcal{W} = \partial_i \mathcal{W} + \mathcal{W} \partial_i \mathcal{K}\,.
\end{align}
In a real formulation for the scalars, with real field space metric $G_{ij}$, we find that the function $T = |e^{{\cal K}/2}\mathcal{W}|$
obeys
\begin{equation}
V = G^{ij}\partial_i T\partial_j T - 3 T^2\,.
\end{equation}
If we then perform the following rescalings (change of conventions) $
V \rightarrow  \tfrac{1}{2}V,  G_{ij}\rightarrow \tfrac{1}{4}G_{ij}$
we indeed find that $T$ is the real superpotential $W$ in the conventions (\ref{Bergshoeffconvention1}, \ref{WV}) up to a factor:
$T = \frac{1}{\sqrt8}W$ .

The complex superpotential and the real K\"ahler potential are given by \cite{Giddings:2001yu}
\begin{align}
& \mathcal{W} = \int_{V_6} \Omega \wedge G\,,\\
& \mathcal{K} = -\ln[i(\tau - \bar{\tau})] -3 \ln[-i(\rho-\bar{\rho})]\,,
\end{align}
where the integral is over the six-dimensional internal manifold, $\Omega$ is the complex holomorphic 3-form and $G$ is the complex 3-form $G = F_3 -\tau H_3$, where $ \tau = C_0 + i\e^{-\phi}$ and $\rho = b + i\e^{-4\alpha v/3}$. The axions $b$ and $C_0$ are put to zero. The geometry can be characterized by the following choice of K\"ahler form $J$ and holomorphic 3-form $\Omega = \Omega_R + i\Omega_I$ \footnote{We use the following conventions: $
\Omega\wedge\bar{\Omega}= \frac{4i}{3}J\wedge J\wedge J = 8i \,\text{vol}_6$.}:
\begin{align}
 \Omega_R =& \e^{-3\beta v/2} [\e^{-3\gamma \chi/2} \d y^4\wedge \d y^5 \wedge \d y^6 + \e^{\gamma\chi/2}\d y^2\wedge \d y^3 \wedge \d y^6 + \\& \e^{\gamma\chi/2}\d y^1\wedge \d y^3 \wedge \d y^4 + \e^{\gamma\chi/2}\d y^1\wedge \d y^2 \wedge \d y^5] \,,\nonumber\\ 
 \Omega_I =& \e^{-3\beta v/2} [\e^{3\gamma \chi/2} \d y^1\wedge \d y^2 \wedge \d y^3 + \e^{-\gamma\chi/2}\d y^1\wedge \d y^4 \wedge \d y^5 - \\ & \e^{-\gamma\chi/2}\d y^2\wedge \d y^5 \wedge \d y^6 - \e^{-\gamma\chi/2}\d y^3\wedge \d y^4 \wedge \d y^6)] \,,\nonumber\\ 
J=& \e^{-\beta v} [\d y^1 \wedge \d y^6 + \d y^2 \wedge \d y^4 + \d y^3 \wedge \d y^5]\,.
\end{align}
This indeed reproduces the internal metric (\ref{metric4Dreduction}), as can be verified from $g_{mn} =  -I^l_{\,\,m} J_{ln}$ where $I$ is the complex structure:
\begin{equation}
I^{l}_{\,\,k} = c\varepsilon^{m_1m_2m_3m_4m_5 l}(\Omega_R)_{km_1m_2}(\Omega_R)_{m_3m_4m_5}\,,
\end{equation}
with $c$ a constant, such that $I^2=-\mathbbm{1}$. One now readily checks that $T$ coincides with the real superpotential $W$ of equation (\ref{superpot4D}) up to an overall factor.

\providecommand{\href}[2]{#2}\begingroup\raggedright\endgroup

\end{document}